\pdfoutput=1

          [2001/04/25 1.1 (PWD)]
\documentclass[a4paper,twocolumn]{esapub2005} 
\usepackage{times}
\usepackage{natbib}
\usepackage{graphicx}
\usepackage{url}

\title{High Energy Sources Observed with OMC}
\author{D. R\'isquez}
\author{A. Domingo}
\affil{Laboratorio de Astrof\'isica Espacial y F\'isica Fundamental (LAEFF-INTA), P.O.Box 50727, E-28080, Madrid, Spain; risquez@laeff.inta.es, albert@laeff.inta.es}
\author{J.M. Mas-Hesse}
\affil{Centro de Astrobiolog\'ia (CSIC-INTA), P.O.Box 50727, E-28080, Madrid, Spain; mm@laeff.inta.es}
\author{E. Kuulkers}
\affil{ISOC, ESA/ESAC, Urb. Villafranca del Castillo, P.O.Box 50727, E-28080, Madrid, Spain; Erik.Kuulkers@esa.int}

\begin{document}

\keywords{Techniques: photometric; Galaxy: bulge; X-rays: binaries}

\maketitle


\begin{abstract}

The INTEGRAL Optical Monitoring Camera, OMC, has detected many high energy sources. We have obtained V-band fluxes and light curves for their counterparts. In the cases of previously unknown counterparts, we have searched for characteristic variations in optical sources around the high-energy target position. Results about the Galactic Bulge Monitoring, INTEGRAL Gamma-Ray sources (IGR), and other high energy sources are presented.

\end{abstract}


\section{Introduction}

The Optical Monitoring Camera, OMC \citep{Mas-Hesse2003}, observes the optical emission from the prime targets of the gamma-ray instruments on board the ESA mission INTEGRAL: SPI (gamma-ray spectrometer) and IBIS (gamma-ray imager), with the support of the JEM-X monitor in the X-ray domain. OMC has the same field of view (FOV) as the fully coded FOV of JEM-X, and it is co-aligned with the central part of the larger fields of view of IBIS and SPI. This capability provides invaluable diagnostic information on the nature and the physics of the sources over a broad wavelength range.

The OMC is based on a refractive optics with a Johnson V filter passband (centred at $550$~nm). It has an aperture of 50 mm focused onto a large format CCD ($ 1024 \times 2048 $~pixels) working in frame transfer mode ($ 1024 \times 1024 $~pixels imaging area). Its field of view is $5^{\circ} \times 5^{\circ}$ and image scale is $17.5$~arcsec/pixel. The point spread function is a Gaussian with FWHM (Full Width at Half Maximum) $\simeq 1.3$~pixels ($\simeq 24''$). Point source location accuracy is around $2''$ (1 sigma).

It works properly in a $10^4$ flux factor range, from $V \simeq 7$~mag (saturation effects appears) to $V\approx16-17$~mag (magnitude limit for 3 sigma source detection, combining all shots per pointing). This magnitude limit depends on sky background and source contamination, which can be very important. The OMC single shots are variable, currently there is a 10, 50 and 200 seconds exposures cycle.

Due to telemetry restrictions, only a fraction of the CCD image is transmitted to Earth, usually 1\% of its surface. Then, many subwindows (typically 100, and in any case less than 228) of $11 \times 11 $~pixels are extracted and sent to ground. The targets to be monitored have to be preselected on ground. For this purpose, an OMC Input Catalogue \citep{Domingo2003} has been compiled. It contains most gamma and X-ray sources, variable stars, and HIPPARCOS and Tycho reference stars for astrometric and photometric calibration. The Catalogue contains currently more than $500\,000$ targets.

Fluxes are calculated in 3 different nearly circular apertures: 1, 3, and 5 diameter pixels. Absolute photometric calibration is achieved by comparison with a number of reference standard stars within the field of view of the instrument.

All the data processed by OMC is open to the scientific community. The OMC team has developed a scientific archive, containing the data generated by the OMC and an access system capable of performing complex searches, complementary to the INTEGRAL Archive hosted at ISDC. The system is reachable at:

\url{http://sdc.laeff.inta.es/omc}


\section{Optical Counterparts of 2nd IBIS/ISGRI Soft Gamma-Ray Survey Catalog Sources}

The Second IBIS/ISGRI Soft Gamma-Ray Survey Catalog \citep{Bird2006} was obtained with the IBIS/ISGRI gamma-ray imager on board the INTEGRAL satellite. It is based on more than 10 Ms of observations during the first 2 years of Core Program and public observations, and it covers around 50\% of the whole sky.

This catalog comprises 209 high energy sources. 13 of them are new IGRs recently discovered and they are not in the OMC Input Catalog yet. On the other hand, there are 196 objects observed by OMC.

The mean position error for all the sources detected by IBIS/ISGRI with significance above 10 sigma is $\sim 40''$, enough to identify most of them with a known X-ray counterpart, but not enough in optical wavelengths because sometimes there are more than one optical sources inside the error circle.

In table \ref{tab:statistical} we show some statistical results. In 51 cases (26 \%) the counterpart is detected by OMC (catalog magnitude $ V \leq 17 $~mag). For all these cases we obtain useful light curves, except some of them which are contaminated with flux from a close bright source.

If sources are weaker than $V=17$~mag, OMC can give us upper flux limits or detect their flares.

\begin{table}[t]
  \begin{center}
    \caption{2nd IBIS/ISGRI soft gamma-ray survey catalog. OMC statistical results.}\vspace{0.2cm}
    \renewcommand{\arraystretch}{1.2}
    \begin{small}
      \begin{tabular}[h]{lr}
      \hline
      Number of sources from the 2nd IBIS/ISGRI                           &       \\
      Soft Gamma-Ray Survey that are also                                 &       \\
      included in the OMC Input Catalog                                   &  196  \\
      \hline
      Not observed sources by OMC                                         &  10\% \\
      Observed sources by OMC                                             &  90\% \\
      \hspace{0.3cm}There are source(s) detected inside the error circle  &  78\% \\
      \hspace{0.3cm}There isn't a source detected inside the error circle &  12\% \\
      \hline
      Coordinates error ($1 \sigma$) $<0.1$~pix ($\simeq 2''$)            &  60\% \\
      Coordinates error ($1 \sigma$) $<0.5$~pix ($\simeq 9''$)            &  70\% \\
      Coordinates error ($1 \sigma$) $<2.0$~pix ($\simeq 35''$)           &  81\% \\
      \hline
      Known optical counterpart                                           &  39\% \\
      \hspace{0.3cm}$V > 17$~mag (undetectable by OMC)                    &  13\% \\
      \hspace{0.3cm}$V \leq 17$~mag (detectable by OMC)                   &  26\% \\
      \hline \\
      \end{tabular}
    \end{small}
    \label{tab:statistical}
  \end{center}
\end{table}

Source coordinates errors are important for better data processing. OSA 6.0 gives the best results with coordinates errors smaller than $0.1$ pixel ($2''$), and it works properly with errors smaller than 2 pixels ($35''$).


\subsection{High Energy Sources without Known Optical Counterpart}

\begin{figure}[t]
\centering
\includegraphics[width=0.49\linewidth]{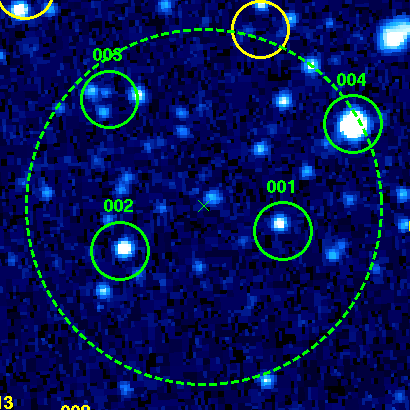}
\includegraphics[width=0.49\linewidth]{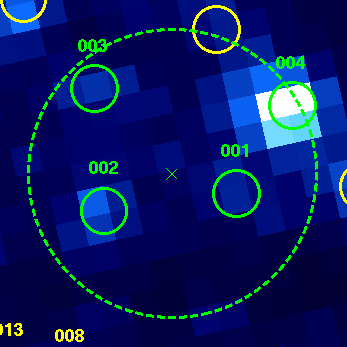}
\\
\vspace{0.4cm}
\includegraphics[width=0.9\linewidth]{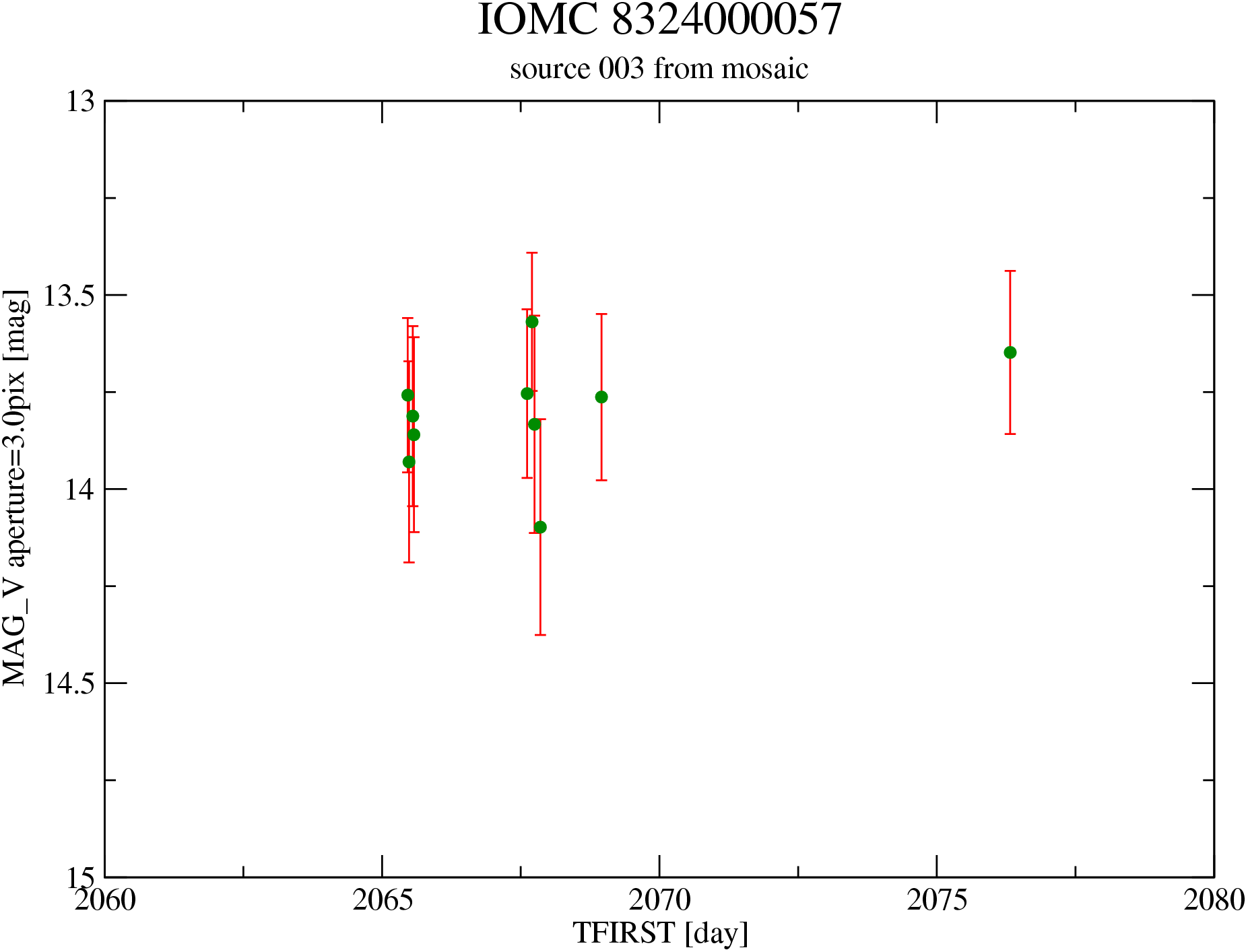}
\caption{Gamma ray source IGR J16207-5129 (IOMC 8324000057). Field view on the left is a DSS image (from ESO Online Digitalised Survey and Space Telescope Science Institute), and field view on the right is an OMC mosaic. The north is up and the east is on the left. The cross at the centre of the field view marks the high energy source coordinates. Dashed circle centred in the cross has a radius equal to 4 sigma coordinates error (at 99.99\% confidence). Small circles are sources detected by OMC. Optical sources have a 3 digit identification number, sorted by distance to the high energy coordinates. The plot below is an example of light curve.\label{fig:withoutKnownCounterpart_8324000057}}
\end{figure}

There are 13 sources with unknown counterpart and coordinates error greater than 5 pixels ($\simeq 1.5'$). In these cases OMC reads mosaics. These mosaics are groups of $11 \times 11$ pixel subwindows, made up of $2 \times 2$, $3 \times 3$, $4 \times 4$, and $5 \times 5$ subwindows.

In these cases there are many optical sources inside the high energy source error circle (sometimes up to hundreds). Using OMC we obtain photometric measures and light curves for all of them. OMC plots could help to disentangle which the real counterpart is.

These mosaics have been processed using SExtractor \citep{Bertin1996} and a hand made pipeline \citep{Risquez2005}. Usually there is one photometric points per shot. Sometimes, while working with weak sources, a greater signal to noise ratio is needed and then all points per pointing ($\approx 30$~minutes) are combined together.

Figure \ref{fig:withoutKnownCounterpart_8324000057} is an example of OMC results. IGR J16207-5129 is a source detected by INTEGRAL for the first time. Light curves for the four objects inside the error circle are plotted. They are almost constant, we do not find any significant magnitude variation. 

In a recent paper, \cite{Tomsick2006} have found that the source is a HMXRB (High Mass X-Ray Binary), and the counterpart is USNO-A2.0 0375-27093111. This counterpart has B=18.9~mag and R=15.6~mag, and it is not detected by OMC.


\subsection{High Energy Sources with Known Optical Counterpart}

\begin{figure}[t]
\centering
\includegraphics[width=0.9\linewidth]{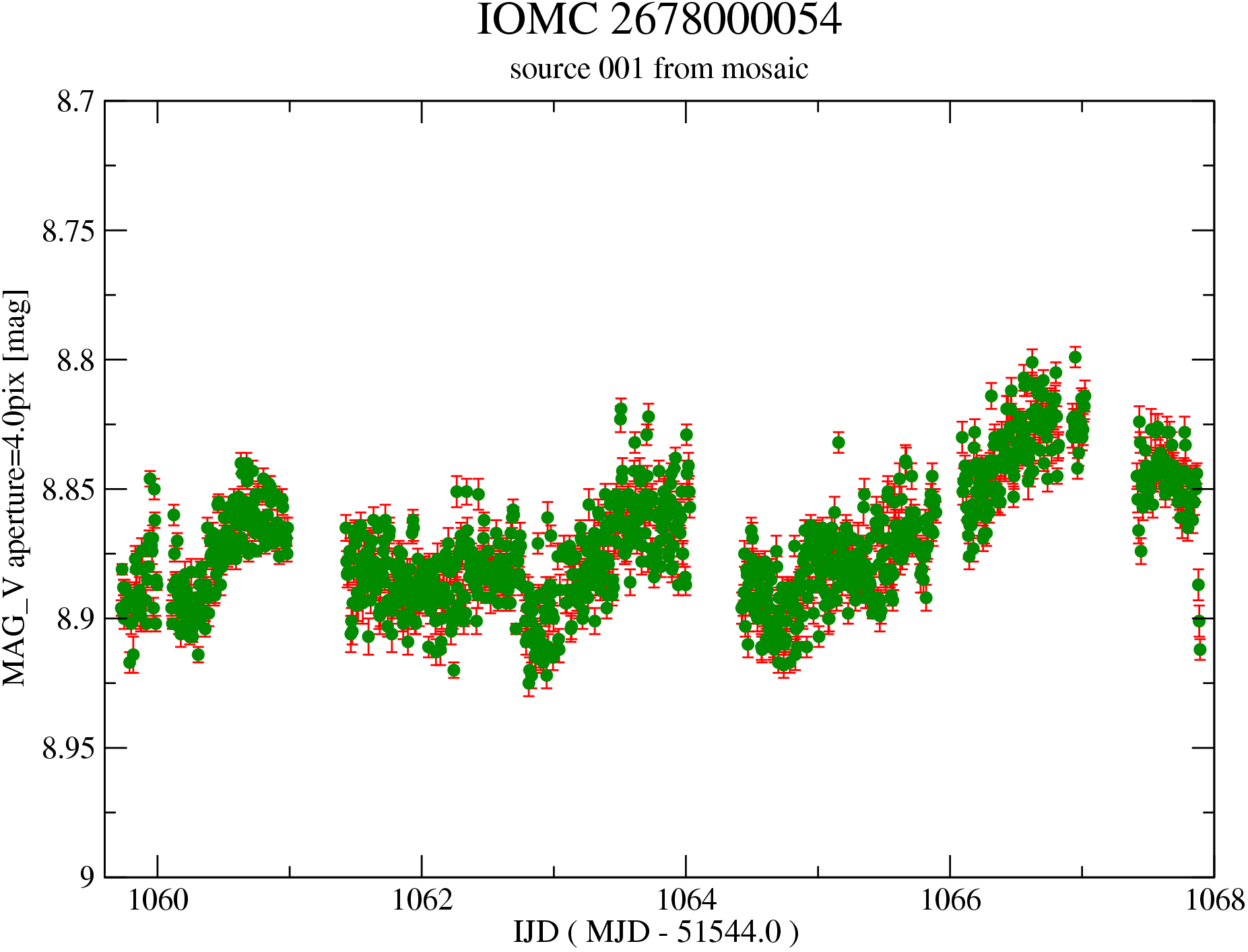}
\\
\vspace{0.4cm}
\includegraphics[width=0.9\linewidth]{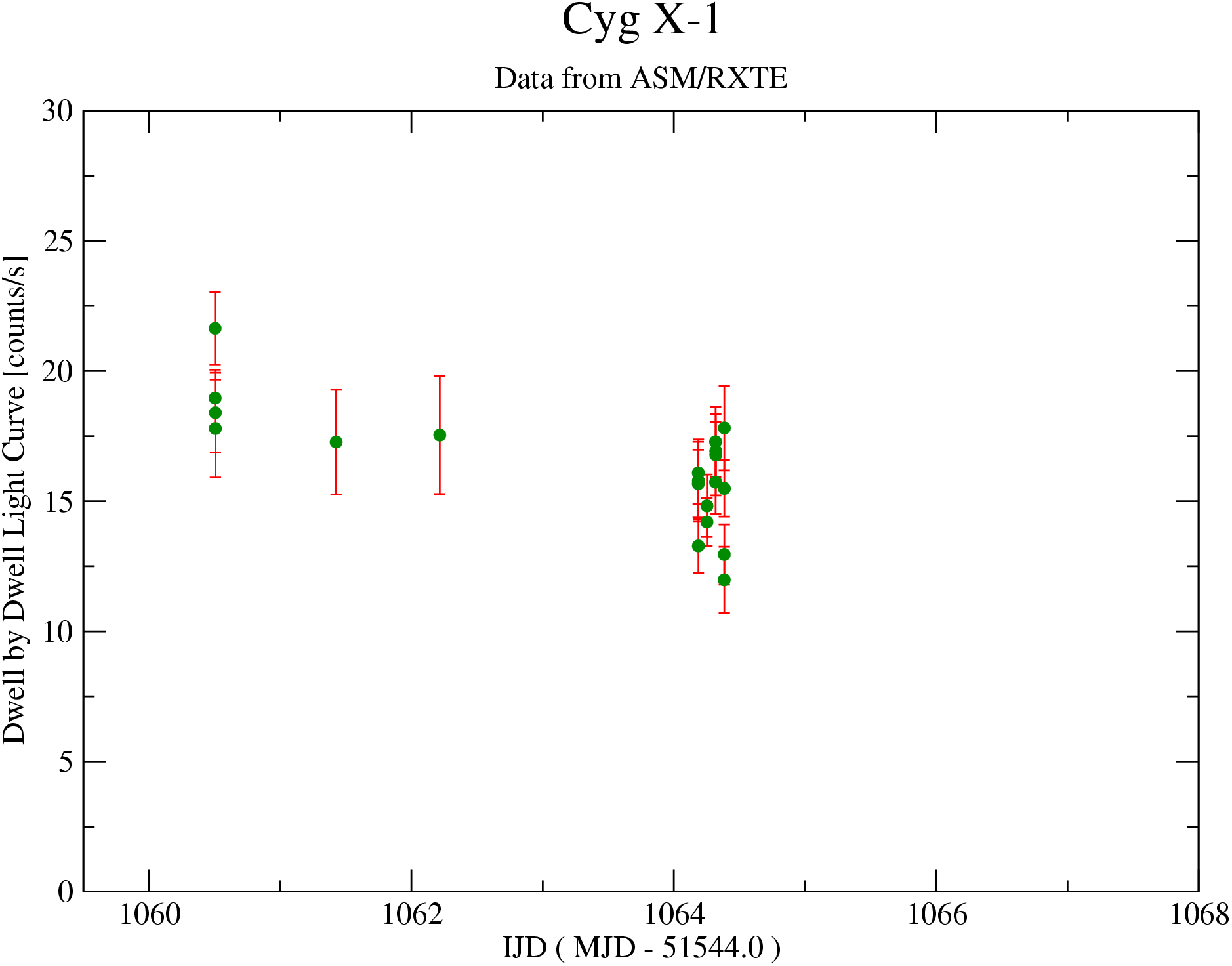}
\caption{High Mass X-Ray Binary Cyg X-1 (IOMC 2678000054). Upper plot is the optical light curve from OMC. Lower plot is the ASM/RXTE light curve (it is in the 2-10~keV range).\label{fig:withKnownCounterpart_CYG_X-1}}
\end{figure}

\begin{figure}[t]
\centering
\includegraphics[width=0.9\linewidth]{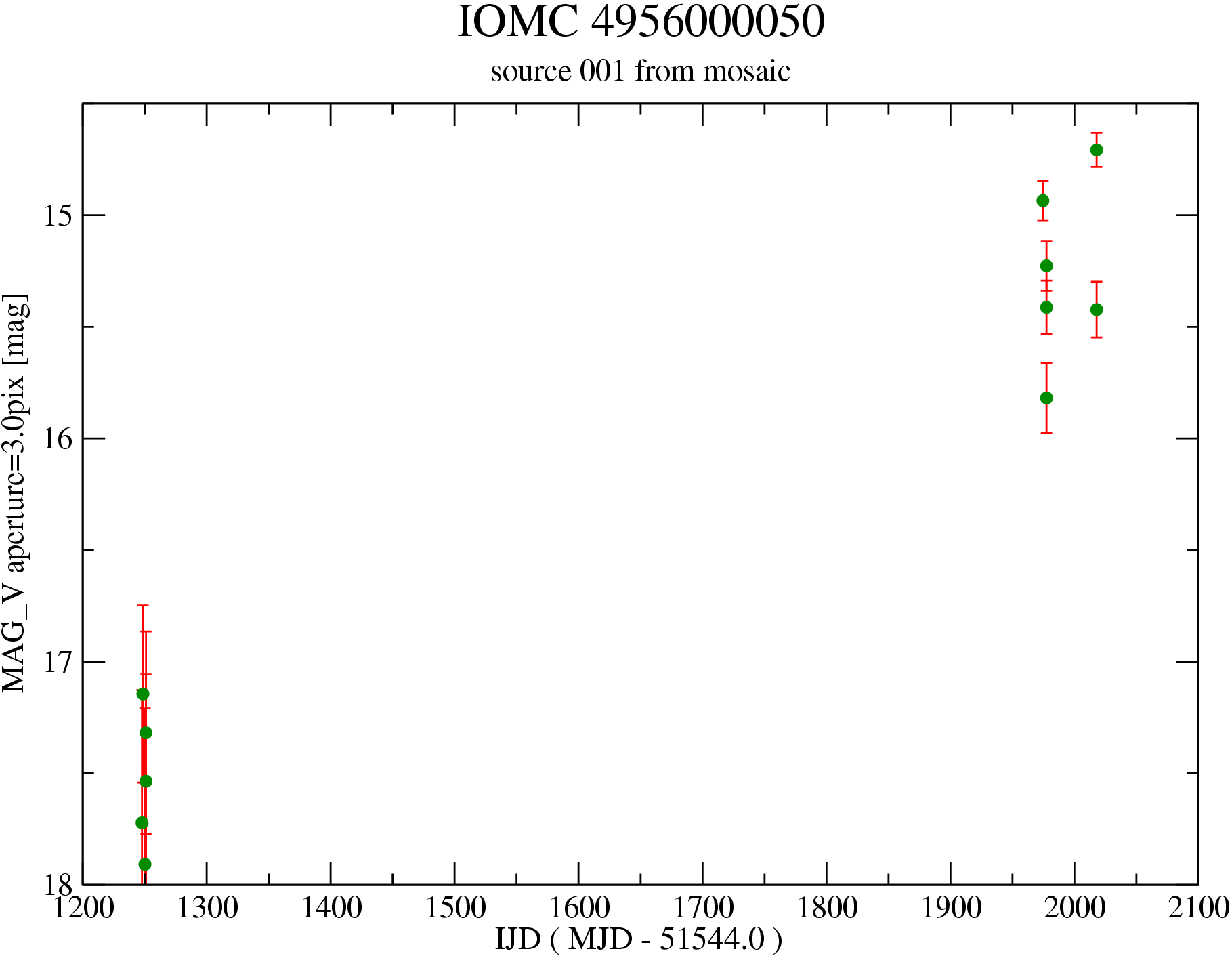}
\\
\vspace{0.4cm}
\includegraphics[width=0.9\linewidth]{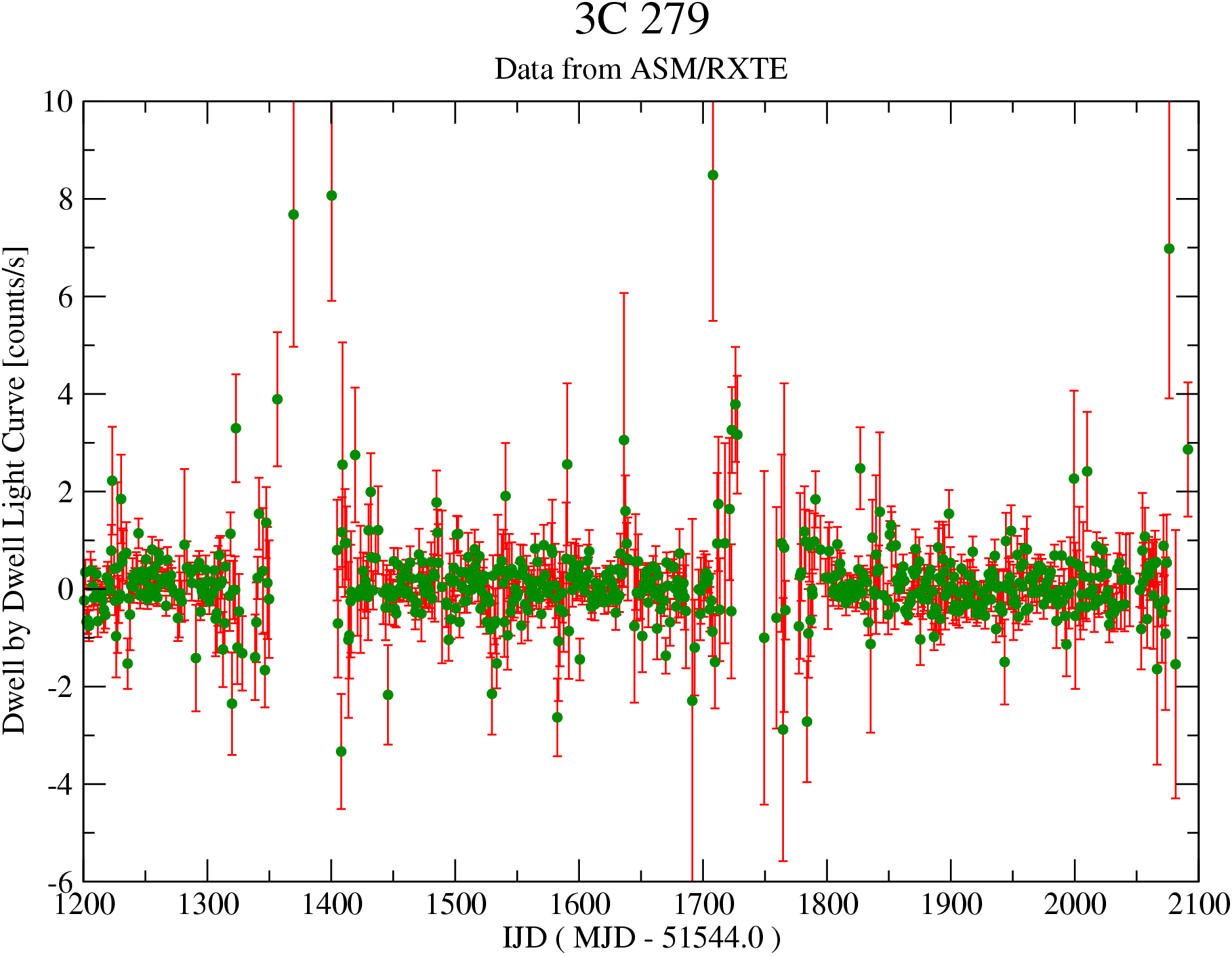}
\caption{Blazar 3C 279 (IOMC 4956000050). Upper plot is the optical light curve from OMC. Lower plot is the ASM/RXTE light curve (it is in the 2-10~keV range).\label{fig:withKnownCounterpart_3C279}}
\end{figure}

OMC obtains optical light curves in the same time range as many high energy instruments. As an example, we include quick-look results provided by the ASM/RXTE team in the same time scale as OMC plots. 

Cyg X-1 is an X-ray binary system (see figure \ref{fig:withKnownCounterpart_CYG_X-1}). Its optical counterpart star is HD 226868 \citep{Bolton1972,Webster1972}, also known as V1357 Cyg. This star is a type O9.7Iab blue giant filling its Roche lobe and transferring mass into the black hole.

We can see a variation in optical Cyg X-1 light curve. After detrending a lineal tendency, we have calculated the best period using a phase dispersion minimization method. Its calculated period is 5.6 days, the well known orbital period of the system. It is due to the blue giant star filling its Roche lobe, because the cross section viewed from the Earth depends on the orbital period around the compact object. OMC provides the precise orbital phase for the study of this object.

This object has a nearby star whose light contaminates its photometry. However, OMC achieves enough precision to catch the 0.06 magnitude amplitude variation.

The X-ray emission is originated around the black hole, and reflects the behaviour of the accretion disk.

In figure \ref{fig:withKnownCounterpart_3C279} there is an outburst detected by OMC in the blazar 3C 279, from $V=17.5$~mag to $V=14.7$~mag. Others optical outbursts (in R filter, from 16.7 to 12.5 magnitude) in the last years are published by \citet{Balonek2002}. Nevertheless, the X-ray plot shows that 3C 279 is not detected.

As other blazars, 3C 279 light curve is expected to show microvariability (around 0.05 magnitudes in a short time scale), flares (around 1 magnitude in one week) and outbursts (many magnitudes in months or years). OMC allows us to analyze these long term variations easily.


\section{Galactic Bulge Monitoring}

The Galactic Bulge is a region rich in bright variable high-energy X-ray and gamma-ray sources. INTEGRAL observes this region regularly during all the visibility periods, in a specific program called Galactic Bulge Monitoring \citep{Kuulkers2006}. The main aim is to investigate the source variability and transient activity on time scales of days to weeks to months at hard X-ray and soft gamma-ray. One complete hexagonal dither pattern (7 pointings of 1800 seconds each) is performed during each INTEGRAL revolution, roughly ever 3 days.

As a service to the scientific community, the JEM-X and IBIS/ISGRI light curves in two energy bands (JEM-X: $3-10$~keV and $10-25$~keV, ISGRI: $20-60$~keV and $60-150$~keV) are made publicly available as soon as the observations are performed in:

\url{http://isdc.unige.ch/Science/BULGE}

OMC monitors all sources detected by IBIS/ISGRI in the Galactic Bulge which are in its field of view. The full analysis is working in an automatic way, producing the results (light curves) in a short period of time, usually less than 24 hours after the observation is performed. At the moment, there are 26 sources with wide light curves and useful data.

Because the Galactic Bulge region is very crowded for OMC, in the flux extraction process we force the photometric aperture to be centred at the source coordinates, which are taken from the OMC Input Catalogue. This is a new functionality in OMC OSA 6.0 (OSA, Off-line Standard Analysis, distributed by ISDC). It allows us to monitor not only those sources already detected by OMC, but also those previously undetected that could show bright flares detectable occasionally by OMC.

For all sources we obtain one photometric point per OMC shot. In addition, for weak sources we combine the shots in one photometric point per each pointing in the hexagonal dithering.

\begin{figure}[t]
\centering
\includegraphics[width=0.9\linewidth]{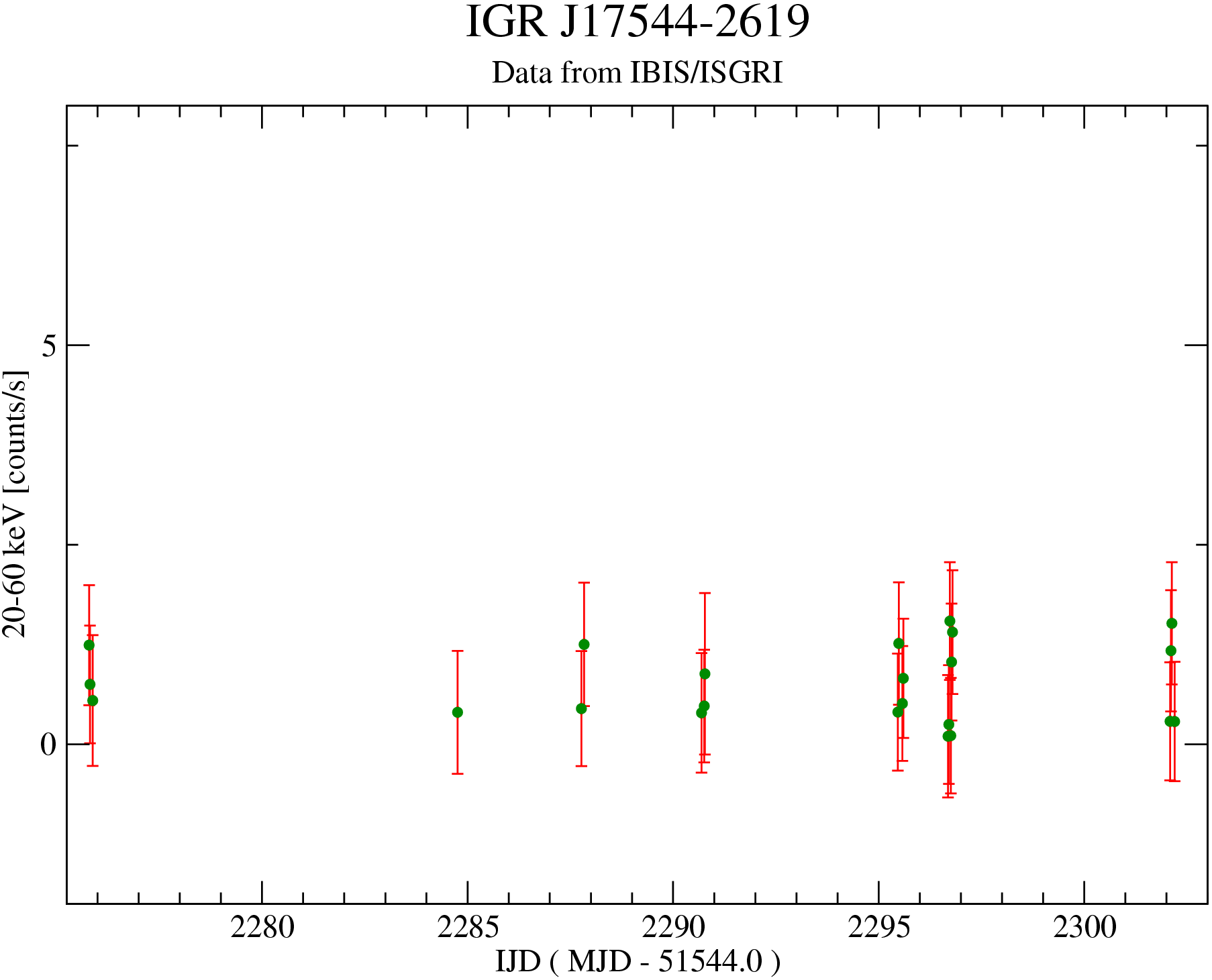}
\\
\vspace{0.4cm}
\includegraphics[width=0.9\linewidth]{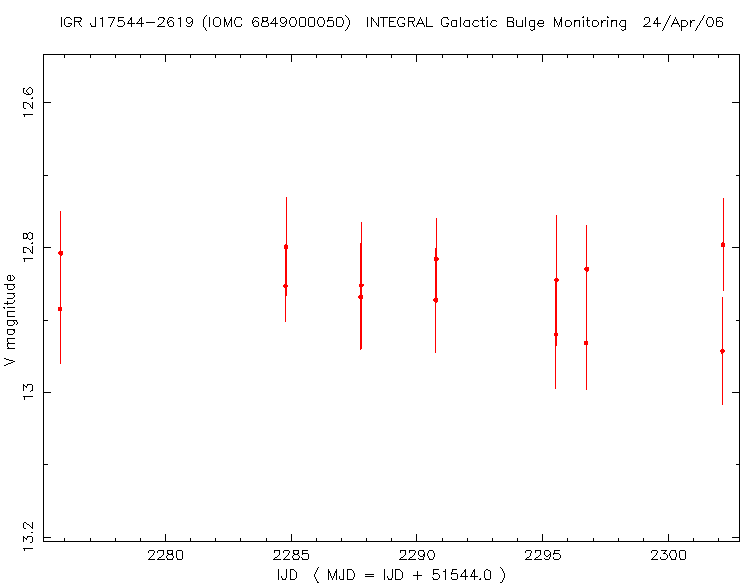}
\caption{Galactic Bulge Monitoring source example: IGR J17544-2619 (IOMC 6849000050). Upper plot: IBIS/ISGRI, lower plot: OMC.\label{fig:IGR_J17544-2619}}
\end{figure}

Figure \ref{fig:IGR_J17544-2619} is an example of simultaneous light curves in two different energy ranges. The source is IGR J17544-2619 (IOMC 6849000050), probably a HMXRB \citep{intZand2005}. With OMC we observe its probable counterpart (2MASS J17542527-2619526, \citet{Rodriguez2003}), although there are other fainter candidates in the 2MASS Archive not detectable by OMC.

Anyway, this source is also in USNO B1.0 catalog. Its magnitudes are B2=13.0 and R2=12.0. OMC detects this source with a constant magnitude V=12.9.

\section*{Acknowledgments}

This research is partially suported by spanish MEC under grant ESP2005-07714-C03-03.

This research has made use of the SIMBAD database, operated at CDS, Strasbourg, France.



\begin{thebibliography}{}


\bibitem[Mas-Hesse et al.(2003)]{Mas-Hesse2003}
Mas-Hesse J.~M., Gim\'enez A., Domingo A., et al.\ 2003, A\&A, 411, L261

\bibitem[Domingo et al.(2003)]{Domingo2003}
Domingo, A., et al.\ 2003, A\&A, 411, L281 

\bibitem[Bird et al.(2006)]{Bird2006}
Bird, A.~J., et al.\ 2006, ApJ, 636, 765 

\bibitem[Bertin \& Arnouts(1996)]{Bertin1996}
Bertin, E., \& Arnouts, S.\ 1996, A\&AS, 117, 393 

\bibitem[R\'isquez(2005)]{Risquez2005}
R\'isquez D.\ 2005, {\it Fotometr\'ia \'Optica desde el espacio}, Master Thesis

\bibitem[Tomsick et al.(2006)]{Tomsick2006}
Tomsick, J.~A., Chaty, S., Rodriguez, J., Foschini, L., Walter, R., \& Kaaret, P.\ 2006, ApJ, 647, 1309

\bibitem[Bolton(1972)]{Bolton1972}
Bolton, C.~T.\ 1972, Nat, 240, 124

\bibitem[Webster and Murdin(1972)]{Webster1972}
Webster B.~L. \& Murdin P.\ 1972, Nat, 237, 507

\bibitem[Balonek \& Kartaltepe(2002)]{Balonek2002}
Balonek, T.~J., \& Kartaltepe, J.~S.\ 2002, {\it Bulletin of the American Astronomical Society}, 34, 1109 

\bibitem[Kuulkers et al.(2006)]{Kuulkers2006}
Kuulkers E., Shaw S.~E., Brandt S., et al.\ 2006, A\&A in preparation

\bibitem[in't Zand(2005)]{intZand2005}
in't Zand J.~J.~M.\ 2005, A\&A, 441, L1

\bibitem[Rodriguez(2003)]{Rodriguez2003}
Rodriguez J.\ 2003, ATel \#194

\end{thebibliography}
\end{document}